\documentclass[aps,pre,reprint,groupedaddress,longbibliography,floatfix]{revtex4-2}

\usepackage{graphicx} 
\usepackage{dcolumn}
\usepackage{amsmath,amssymb}
\usepackage{physics}
\usepackage[dvipsnames]{xcolor}
\usepackage{times, mathptmx}
\usepackage{bm}
\usepackage[T1]{fontenc}
\usepackage{placeins}
\usepackage{changepage}
\usepackage{threeparttable}
\usepackage{enumitem}
\usepackage{nameref}
\usepackage{xr}
\usepackage[colorlinks=true, linkcolor=black, citecolor=black, urlcolor=black]{hyperref}
\hypersetup{filecolor=black}

\usepackage{lmodern}
\usepackage{xcolor}
\usepackage{etoolbox}
\usepackage{changes}
\setaddedmarkup{\textcolor{black}{#1}}
\setdeletedmarkup{}

\newcommand{\muh}{$\mu_{\mathrm o}$}
\newcommand{\mua}{$\mu_1$}
\newcommand{\ru}{${\mathrm{R_{o}}}$}
\newcommand{\rd}{$\mathrm{R_1}$}

\newcommand{\na}{$n_1$}
\newcommand{\nh}{$n_{\mathrm o}$}

\newcommand{\ph}{$\mathcal{P}_{\mathrm o}$}

\begin{document}

\title{Two faces of symmetry match in gear-like coupled rotary motors}

\author{Sara Iranbakhsh}
\email{sara.iranbakhsh@ista.ac.at}
\altaffiliation[Current address: ]{Institute of Science and Technology Austria (ISTA), Klosterneuburg, Austria}
\author{David A.\ Sivak}
\email{dsivak@sfu.ca}

\affiliation{Department of Physics, Simon Fraser University, Burnaby, British Columbia V5A 1S6, Canada}

\date{\today}

\begin{abstract}
Rotary molecular motors commonly couple components with different rotational symmetries, but how this gearing controls energy transduction remains unclear. Here we show that symmetry can produce a fundamental tradeoff in coupled stochastic barrierless rotary motors: the gear ratio that maximizes output power under tight coupling is also the ratio that requires the strongest coupling to remain locked. Consequently, weakening the coupling can invert the effect of gearing, transforming a power maximum into a local minimum. At strong driving, both the power-optimal and most slip-susceptible ratios approach symmetry match. Numerical solutions of the Fokker–Planck dynamics for rotors with finite energy barriers reproduce these predictions and reveal how the behavior changes when driving strength is fixed per rotation or per subunit. Motivated by the bacterial flagellar motor, these results identify coupling strength as a key determinant of whether matched rotational symmetries enhance or suppress energy transduction, providing a general design principle for coupled molecular motors.
\end{abstract}

\maketitle

\section{Introduction}
Biological molecular motors are isothermal, cyclic machines operating at energy scales comparable to the thermal energy $k_{\rm B}T$~\cite{kuhlbrandt_structure_2019}. Constantly perturbed by collisions with water or other proteins, their components experience large stochastic fluctuations~\cite{brown_theory_2020}, and with Reynolds numbers on the order of $10^{-8}$~\cite{holzwarth_forces_2002}, they operate in the overdamped regime. Consequently, the key to understanding molecular motors lies in nonequilibrium and statistical approaches applied to isothermal, overdamped, and thermodynamically consistent models. 

\begin{figure*}[t!]
\centering  
\includegraphics[width=\textwidth]{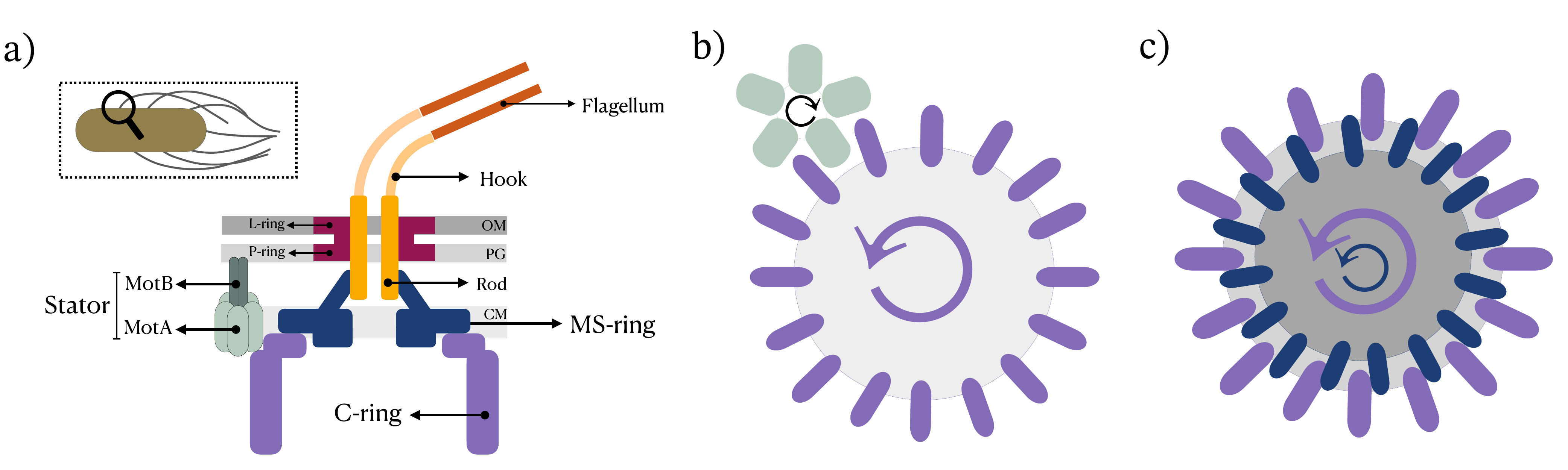}
    \caption{Schematics of \emph{Salmonella}'s flagellar motor structure. a) The BFM consists of multiple rings: the C-ring, MS-ring, P-ring, and L-ring. A stator (composed of MotA and MotB proteins) binds to the peptidoglycan (PG) layer and rotates driven by ion flow. Rotation of the stators rotates the C-ring, which rotates the MS-ring, which in turn rotates the rod and the hook, and hence the flagellum. The motor is embedded in multiple layers: outer membrane (OM), PG, and cytoplasmic membrane (CM). b) When stators (light green) interact with the C-ring (purple), they drive the ring to rotate. c) Rotation of the C-ring drives rotation of the MS-ring (blue) in the same direction.}
    \label{fig:BFM}
\end{figure*}

The vast majority of living organisms depend on rotary molecular motors~\cite{yoshida_atp_2001}. Among the distinct ion-driven rotary motors (also called rotors) reported in nature are ATP synthase and the bacterial flagellar motor (BFM)~\cite{junge_torque_2009,takekawa_structure_2020,nirody_atp_2020}. The BFM is one of the most complex biological rotary motors and is located at the base of each flagellum~\cite{nord_mechanisms_2020}. Bacterial motility is enabled by stators that utilize the free energy of ion flow to apply torque to rotate the rotor and, consequently, the flagellum~\cite{reid_maximum_2006} (Fig.~\ref{fig:BFM}). Each stator consists of five MotA and two MotB proteins in proton-driven rotors, or their PomA and PomB counterparts in sodium-driven rotors~\cite{deme_structures_2020,homma_periplasmic_2022}. The stators directly contact the C-ring~\cite{terashima_site-directed_2021}, whose upper surface is attached to the bottom of the MS-ring (Fig.~\ref{fig:BFM}).

Symmetry mismatch is also observed in other components of the BFM~\cite{nakamura_structure_2024}, where it is proposed to enhance flexibility and adaptability during assembly~\cite{kawamoto_native_2021,vartanian_structure_2012}. Surprisingly, the symmetries of BFM components vary not only across species but also among individual motors within the same species~\cite{carroll_flagellar_2020,chang_molecular_2020}. The native \emph{Salmonella} consistently has 34 subunits (FliF) on the MS-ring, while the C-ring subunits (FliG) show variation in their symmetry (32--36-fold), with 34-fold being the most common~\cite{kawamoto_native_2021, johnson_symmetry_2020}. Since the C-ring size determines the maximum number of stators that can be accommodated~\cite{nakamura_flagella-driven_2019}, larger C-rings enable higher torque generation through interaction with more stators. Thus, the number of C-ring subunits appears to have evolved to suit the specific needs of each bacterium~\cite{chang_structural_2021}; for example, bacteria that inhabit highly viscous environments (where the load on the BFM is larger) have been observed to have larger C-rings~\cite{drobnic_situ_2025}. 

In protein complexes, symmetry mismatch has already been proposed to enable functionality by permitting mechanisms such as relative motion between components of differing symmetry~\cite{goodsell_structural_2000}, for example, facilitating DNA packaging in bacteriophage~\cite{hendrix_symmetry_1978}. Recent theoretical work has shown that symmetry mismatch affects the dynamics and performance of arbitrary (spring-like) coupled rotary motors modeled on the $\rm F_{o}$ and $\rm F_1$ dynamics in ATP synthase~\cite{iranbakhsh_effects_2025}, while other studies have investigated how such mismatch is mechanically accommodated during rotation in ATP synthase~\cite{kubo_fo-f1_2023}. However, neither aspect has been investigated in coupled rotary motors resembling the BFM, with direct interactions unmediated by any intervening parts (dubbed gear-like coupling). A deeper understanding of such gear-like coupled systems could help answer fundamental questions about how evolution may tune the BFM's symmetries to improve its function under varying conditions, inform the design of synthetic rotary molecular motors~\cite{pogoryelov_engineering_2012}, and guide structure-based drug development targeting the BFM~\cite{andries_diarylquinoline_2005,preiss_structure_2015}.

Here, we numerically simulate two rotors coupled through a gear-like coupling under either constant or scaling driving forces. We investigate the output power as a function of the upstream rotor's subunit number and find that under constant driving, with a strong driving force on the upstream rotor, the coupling strength determines whether the output power exhibits a maximum or a local minimum at symmetry match. This strongly affects whether operating at symmetry match is beneficial when high output power is desired. Under scaling driving, we observe no such dependence on coupling strength at symmetry match.

\section{Simple model}
\begin{figure}[htb]
    \centering
    \includegraphics[width=\linewidth]{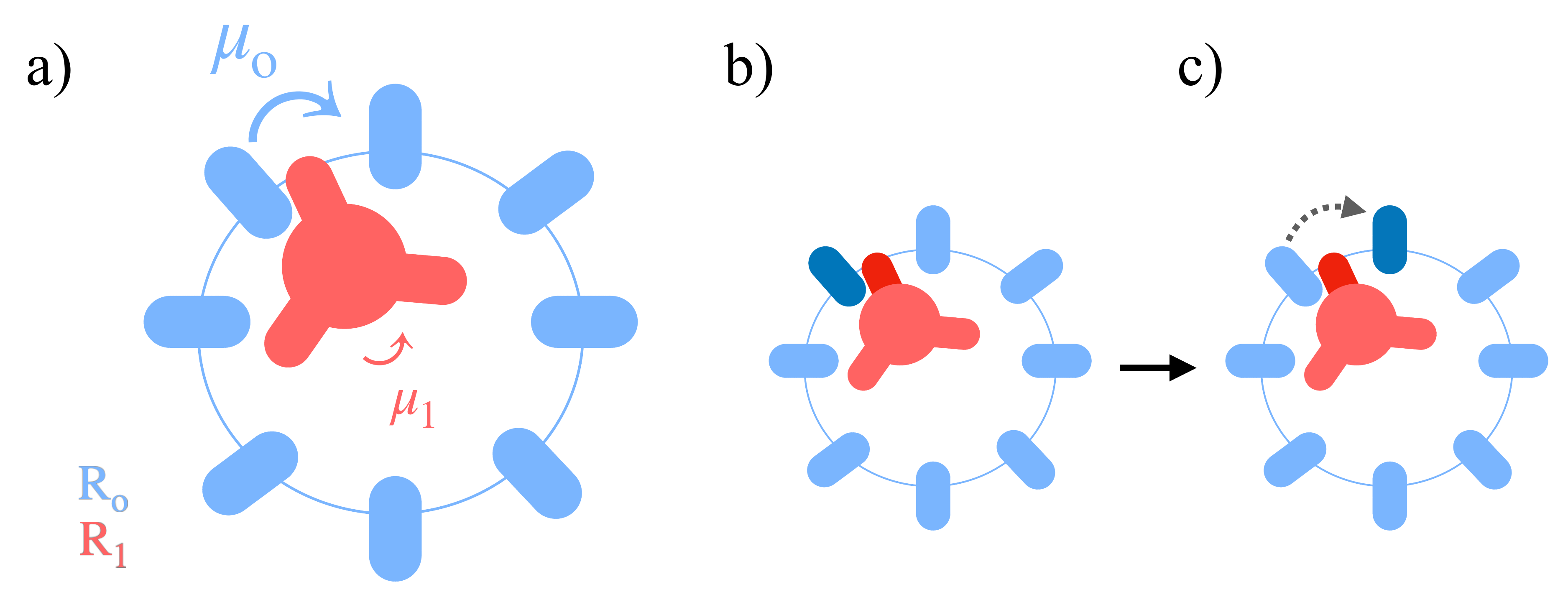}
    \caption{a) Schematic of gear-like coupling between arbitrary rotors \ru\ (blue) and \rd\ (red). The number of teeth on each gear corresponds to the number $n$ of subunits. \muh\ and \mua\ represent the respective driving forces (constant or scaling) on \ru\ and \rd. b, c) Schematic of a possible trajectory contributing to slippage under intermediate-strength gear-like coupling. When one tooth from each rotor (here, dark blue and dark red) come in contact (b), the upstream rotor (\ru) can capitalize on the flexible coupling and rotate $2\pi/{n_{\rm o}}$ while \rd\ maintains the same orientation (c).}
    \label{fig:schematic_rotors}
\end{figure}

We model each rotor on its own as a Brownian particle diffusing on an angular energy landscape with $n$ periodic barriers proportional to $\cos(n\theta)$, for angular orientation $\theta$. In a coupled system, two rotors are connected and transduce energy through a coupling potential. In systems such as ATP synthase, where elastic compliance allows relative twisting between the rotors~\cite{sielaff_domain_2008,wachter_two_2011}, a spring-like coupling (which could be thought of as a function of the angular difference between the rotors' orientations) is appropriate~\cite{iranbakhsh_effects_2025}. In the BFM, however, the components have direct interactions that resemble gears in contact, motivating a gear-like coupling that is a function of the weighted difference between the angular orientations, where the weights are the subunit numbers. Denoting the two coupled molecular rotors by \ru\ and \rd, the total energy landscape is
\begin{align}
    V(\theta_{\rm {o}},\theta_1) &= -\tfrac{1}{2}E_{\rm {o}}\cos(n_{\rm {o}} 
    \theta_{\rm {o}}) -\tfrac{1}{2}E_1\cos(n_1 \theta_1)\notag \\
    &\quad -\tfrac{1}{2}E_{\rm c}\cos(n_{\rm {o}} \theta_{\rm {o}}- n_1 \theta_1),
    \label{eq:energy_land_eq_two}
\end{align}
with respective barrier heights $E_{\rm o}$ and $E_1$ along \ru's and \rd's potentials, and coupling strength $E_{\rm c}$. Figure~\ref{fig:schematic_rotors}a shows a schematic of gear-like coupling. 

Following previous studies~\cite{iranbakhsh_effects_2025}, we consider two driving schemes motivated by different interactions in the BFM. Constant driving, corresponding to a fixed free-energy change per full rotation, is appropriate for the C-ring--MS-ring interaction, where the driving originates from the stators and is independent of the C-ring's subunit number. Scaling driving, corresponding to a fixed free-energy change per subunit, is motivated by the stator--C-ring interaction, where each stator subunit translocates, on average, two ions per rotation, giving a driving force proportional to the number of subunits:
\begin{equation}
    \mu_{\rm o/1} = n_{\rm o/1} \times \mu_{\rm o/1}^{\rm sub}.
\end{equation}

\begin{figure*}[t!]
    \centering
    \includegraphics[width=\textwidth]{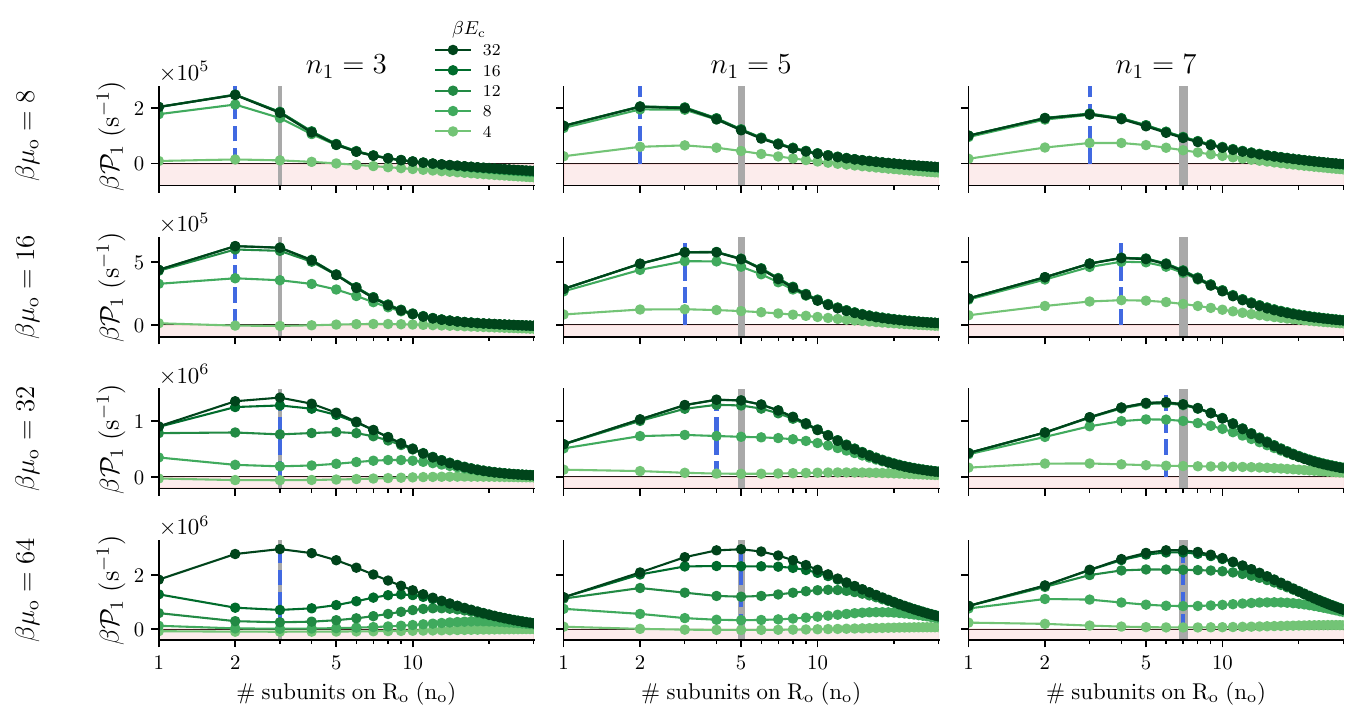}
    \caption{Output power $\beta \mathcal{P}_{1}$ as a function of the number \nh\ of \ru\ subunits, for different coupling strengths $\beta E_{\rm c}$ under constant driving. Rows each have fixed driving force $\beta\mu_{\rm o}$ on \ru\ ranging from 8 to 64; columns each have fixed number $n_1$ of \rd\ subunits ranging from 3 to 7. Gray vertical solid lines: symmetry match ($n_1 = n_{\rm o}$). Blue dashed vertical lines: power-maximizing $n_{\rm o}^{*}$ predicted by finite-barrier tight-coupling theory~\eqref{eq:n0_star_tight_barriers_supp}. $\beta \mu_1 = -2$ and $\beta E_{\rm o} = \beta E_1 = 2$ throughout. Here and throughout other figures, shared axes use the same scales, and curves connecting data points are included as guides to the eye.}
    \label{fig:P1_vs_n0_mod3}
\end{figure*}

The dynamics of the coupled system is described by a joint probability distribution over the angular orientations $\theta_{\rm o}$ and $\theta_1$, which evolves according to the Fokker--Planck equation
\begin{align}
\label{eq:FPE}
    &\frac{\partial P(\theta_{\rm o},\theta_{1},t)}{\partial t} =
    \frac{1}{\gamma_{\rm o}}
    \left[\frac{\partial}{\partial \theta_{\rm o}}\left( \frac{\partial V} {\partial \theta_{\rm o}} - \mu_{\rm o} \right)
    +\frac{1}{\beta}\frac{\partial^2}{\partial \theta_{\rm o}^2}\right]P(\theta_{\rm o},\theta_{1},t) \nonumber \\
    &\quad\quad + \frac{1}{\gamma_1}
    \left[\frac{\partial}{\partial \theta_{1}}\left( \frac{\partial V}
    {\partial \theta_{1}} - \mu_1 \right) +\frac{1}{\beta}\frac{\partial^2}{\partial \theta_{1}^2}\right] P(\theta_{\rm o},\theta_{1},t),
\end{align}
for respective driving forces $\mu_{\rm o}$ and $\mu_1$ (with opposite signs) and respective friction coefficients $\gamma_{\rm o}$ and $\gamma_1$ of \ru\ and \rd, inverse temperature $\beta \equiv (k_{\rm B}T)^{-1}$, and total potential landscape $V$~[Eq.~\eqref{eq:energy_land_eq_two}]. We use equal friction coefficients $\gamma_{\rm o}=\gamma_1$, although subunit-dependent friction coefficients may be appropriate for some physical systems. We numerically simulate \eqref{eq:FPE} using a finite-difference scheme in time and space~\cite{githubver3,lathouwers_nonequilibrium_2020,githubEmma2}. At the steady-state distribution $P_{\rm ss}(\theta_{\rm o},\theta_1)$, the local probability fluxes of rotors \ru\ and \rd\ are used to calculate the rotors' fluxes $J_{\rm o/1}$. The input power $\mathcal{P}_{\rm o}$ and output power $\mathcal{P}_1$ are~\cite{lathouwers_internal_2022,ehrich_energy_2023}
\begin{subequations}
\begin{align}
    \mathcal{P}_{\rm o} &= 2\pi \mu_{\rm o} J_{\rm o},\label{eq:input} \\
    \mathcal{P}_{1} &= -2\pi \mu_1 J_{1}, 
    \label{eq:output}
\end{align}
\end{subequations}
where under this sign convention both $\mathcal{P}_{\rm o}$ and $\mathcal{P}_1$ are positive when \ru\ and \rd\ rotate in the direction of \muh, corresponding to a system in which \ru\ consumes energy while \rd\ performs useful work. 

\section{Constant driving}
\label{sec:constant_driving}
Contrary to spring-like coupling where \small{$ V_{\rm c}^{\rm spring} \propto \cos(\theta_{\rm o} - \theta_{1})$}, under gear-like coupling (\small{$V_{\rm c} \propto \cos(n_{\rm o}\theta_{\rm o} - n_1 \theta_{1})$}): even when the individual rotors have flat energy landscapes (i.e., are barrierless), the number of subunits affects the dynamics through the coupling potential. 

Under gear-like coupling, \emph{total driving} is defined to encapsulate the competition between per-subunit free-energy changes:
\begin{equation}
    \mu_{\rm tot}  \equiv \frac{\mu_{\rm o}}{n_{\rm o}} + \frac{ \mu_1 }{n_1}. 
\label{eq:total_driving_gears}
\end{equation}
In the tight-coupling limit---where the two rotors rotate in lockstep---the sign of the total driving denotes the direction the composite system of the two rotors rotates in to reduce environmental free energy, and thus determines the identities of the upstream and downstream rotors: when $\mu_{\rm tot} > 0$, \ru\ acts as the upstream rotor, and both input power and output power are positive (\ru\ consumes energy and \rd\ does work). However, increasing \nh\ decreases $\mu_{\rm o}/n_{\rm o}$, which eventually makes $\mu_{\rm tot} < 0$ and hence \rd\ becomes the upstream rotor, turning both input and output power negative (\rd\ consumes energy and \ru\ does work). Thus, under tight coupling, the threshold number of subunits at which the output becomes negative satisfies $\mu_{\rm tot} = 0$. 

\begin{figure*}[htbp] 
    \centering
    \includegraphics[width=\textwidth]{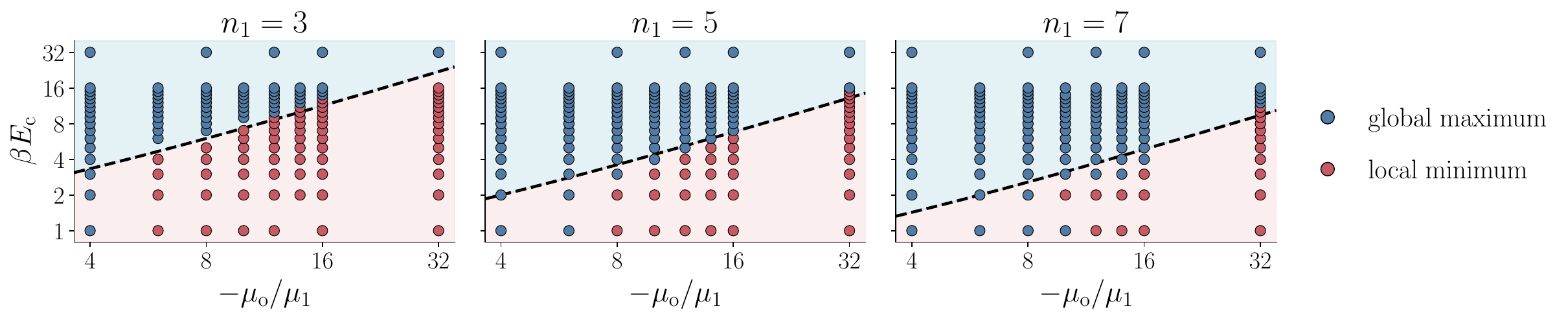}
    \caption{Occurrence of output-power global maxima and local minima under constant driving, as functions of the driving-force ratio $-\mu_{\rm o}/\mu_1$ and coupling strength $\beta E_{\rm c}$, for $n_1=3,5,$ and $7$. For each parameter combination, blue and red circles indicate the presence of an interior global maximum or local minimum, respectively, in $\mathcal{P}_1$ as a function of upstream symmetry $n_{\rm o}$. Extrema at the boundaries ($n_{\rm o}=1,30$) are excluded. Black dashed lines show the deterministic barrierless locking threshold~\eqref{eq:locking_condition_main} evaluated at $n_{\rm o}=n_1$. $\beta\mu_1=-2$ and $\beta E_{\rm o}=\beta E_1=2$ throughout.}
    \label{fig:dip_peak_phase_diagram}
\end{figure*}

\subsection{Strong coupling: driving-dependent optimal \texorpdfstring{\nh}{n_o}}
\label{sec:optimal_n0}

At sufficiently strong coupling, output power exhibits a maximum either at symmetry match or at a slightly smaller number of upstream subunits, $n_{\rm o}^{*}\lesssim n_1$ (Fig.~\ref{fig:P1_vs_n0_mod3}). The location of this maximum depends on the strength of the upstream driving: at \muh\ comparable to $-\mu_1$, output power is maximized at slight symmetry mismatch, whereas increasing \muh\ shifts $n_{\rm o}^{*}$ toward symmetry match, and at sufficiently large \muh\ the maximum occurs at $n_{\rm o}=n_1$.

This behavior can be understood by reducing the tightly coupled system to a single effective coordinate, where the resulting one-dimensional tilted periodic potential gives an analytical expression for the output power (Sec.~\ref{app:tight_barriers}); evaluating this expression at integer $n_{\rm o}$ and selecting its maximum gives the power-maximizing subunit number $n_{\rm o}^*$ shown by the blue dashed lines in Fig.~\ref{fig:P1_vs_n0_mod3}.

The shift of the power maximum toward symmetry match as \muh\ increases can also be shown analytically in the barrierless limit (Sec.~\ref{app:tight_barrierless}). Defining the gear ratio $\kappa\equiv n_1/n_{\rm o}$, the tight-coupling power-maximizing ratio in the barrierless system is:
\begin{equation}  
    \kappa^{*}
    = -\frac{\mu_1}{\mu_{\rm o}}+\sqrt{1+\left(\frac{\mu_1}{\mu_{\rm o}}\right)^2}.
    \label{eq:kappa_star}
\end{equation}
Since $\mu_1<0$, this gives $\kappa^{*}>1$ and therefore $n_{\rm o}^{*}<n_1$ at finite $\mu_{\rm o}$. As $-\mu_{\rm o}/\mu_1$ increases, $\kappa^{*}\rightarrow1$, so that $n_{\rm o}^{*}\rightarrow n_1$. 

This progression toward symmetry match is also seen in simulations more closely motivated by the C-ring--MS-ring interaction, in which the downstream rotor (MS-ring) is passive and differences in the sizes, molecular composition, and local environments of the C-ring and MS-ring may give rise to unequal friction coefficients (Sec.~\ref{SM:C_S_ring}). 

\subsection{Strong driving: local power minimum at symmetry match}
\label{sec:locking_dip}

As \muh\ increases, at strong coupling, the optimum approaches the symmetry match ($n_{\rm o}^{*} \to n_1$), consistent with the $\mu_{\rm o} \gg -\mu_1$ limit of~\eqref{eq:kappa_star}. At weak and intermediate coupling, however, the symmetry-match peak in output power observed at tight coupling flips to a local minimum (Fig.~\ref{fig:P1_vs_n0_mod3}). Symmetry match thus changes from the most favorable (power-maximizing) setup to the (locally) least favorable as the coupling is weakened.

This change is consistent with a crossover from predominantly coordinated rotation to increased relative motion between the rotors: At strong coupling, the rotors move in a coordinated fashion, and their average speeds maintain the fixed gear ratio $\langle\partial_t\theta_{\rm o}\rangle \approx \kappa\langle\partial_t\theta_1\rangle$. At weaker coupling and sufficiently strong driving, \ru\ can instead advance without dragging \rd\ along, despite the coupling interaction. The resulting relative motion constitutes slippage (Figs.~\ref{fig:schematic_rotors}b,c) and reduces the transmission of the upstream driving to the downstream rotor, reducing $J_1$.

In Sec.~\ref{app:barrierless_locking} we show that in a barrierless system, maintaining a locked relative angle requires
\begin{equation}
    E_{\rm c} \geq E_{\rm c}^{\rm lock} = 2\, \frac{n_{\rm o}\mu_{\rm o}-n_1\mu_1}{n_{\rm o}^{2}+n_1^{2}}.
    \label{eq:locking_condition_main}
\end{equation}
For fixed $\mu_{\rm o}$, $\mu_1$, and $n_1$, $E_{\rm c}^{\rm lock}$ is non-monotonic in $n_{\rm o}$ and (defining $\kappa^{\rm lock} \equiv n_1/n_{\rm o}^{\rm lock}$) is maximized at 
\begin{equation}  
    \kappa^{\rm lock} = \kappa^*.
    \label{eq:locking_location_main}
\end{equation}
As $-\mu_{\rm o}/\mu_1$ becomes large, $\kappa^{\rm lock} \rightarrow 1$, and $n_{\rm o}^{\rm lock} \rightarrow n_1$. Thus, in a barrierless system and for strong upstream driving, maintaining a locked relative position at symmetry match requires a stronger coupling strength than at symmetry-mismatched configurations. 

Furthermore, \eqref{eq:locking_location_main} shows that for barrierless rotors, the gear ratio that maximizes output power when the rotors are locked is also the ratio that requires the strongest coupling to maintain locked movement. At strong driving, this produces the symmetry-match power maximum at tight coupling and a symmetry-match local power minimum at weaker couplings (Fig.~\ref{fig:barrierless_power_difference}).

Figure~\ref{fig:dip_peak_phase_diagram} shows the occurrence of the numerically observed power maximum or local minimum with the deterministic barrierless locking threshold~\eqref{eq:locking_condition_main}. At sufficiently strong driving, near or at symmetry match (see Fig.~\ref{fig:P1_vs_n0_mod3}) there is predominantly a local minimum in power below this threshold, whereas a power maximum is more common above it. This correspondence supports the interpretation that the emergence of a local minimum at symmetry match is associated with increased slippage as the coupling becomes insufficient to maintain coordinated rotation. Since the numerical simulations include nonzero rotor barriers and thermal fluctuations, the analytical locking condition does not define a sharp transition in the full system between power maximum and minimum.

\subsection{Disruption}
\label{subsubsec:disruption_gears}
\begin{figure*}[t!]
    \centering   \includegraphics[width=\textwidth]{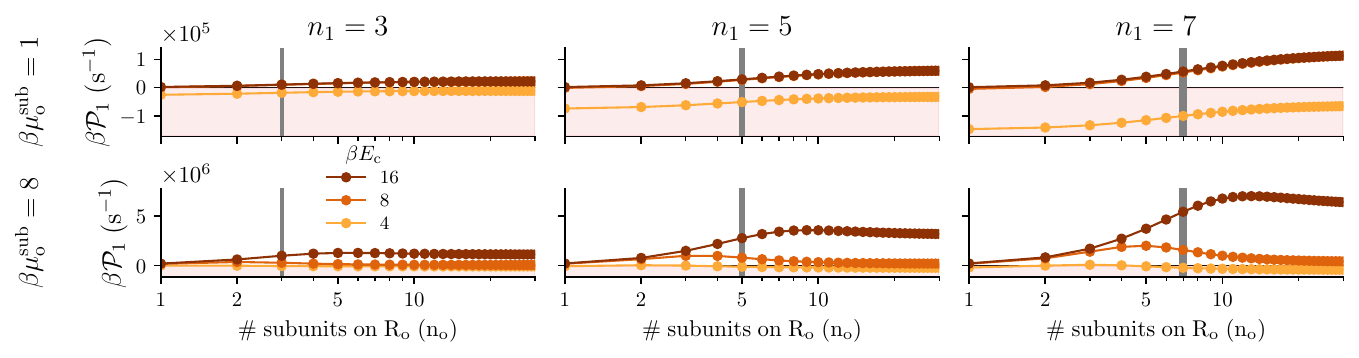}
    \caption{Output power under scaling driving, as a function of the number \nh\ of \ru\ subunits, for different coupling strengths $\beta E_{\rm c}$. Each row has a fixed per-subunit driving force $\beta\mu_{\rm o}^{\rm sub}$ on \ru\ (1 on the top row and 8 on the bottom); each column has fixed number $n_1$ of subunits of \rd\ ranging from 3 to 7. Gray vertical solid lines show symmetry match ($n_1 = n_{\rm o}$). $\beta \mu_1^{\rm sub} = -2/3$ and $\beta E_{\rm o} = \beta E_1 = 2$ throughout.}
    \label{fig:P1_vs_n0_mod4}
\end{figure*}
At fixed intermediate coupling, as \nh\ increases, the output power decreases from positive to negative (Fig.~\ref{fig:P1_vs_n0_mod3}). This change is \emph{disruption}, as reported in~\cite{iranbakhsh_effects_2025}: output power turns negative while input power remains positive throughout (Fig.~\ref{fig:P0_vs_n0_mod3}). For gear-like coupling, disruption only occurs for intermediate-strength coupling (e.g., $\beta E_{\rm c} \approx 4$ for $\beta\mu_{\rm o} = 8$) because under tight coupling, the two rotors are constrained to move with a fixed gear ratio, so \rd\ cannot reverse independently.

Disruption under intermediate coupling happens because increasing \nh\ decreases \ru's per-subunit free-energy change ($\mu_{\rm o}/n_{\rm o}$) to the point that, at the engagement of an \ru\ subunit with an \rd\ subunit, $\mu_{\rm o}/n_{\rm o}$ (as mediated by the coupling) becomes too weak for \ru\ to significantly affect \rd: the influence of \ru\ is outweighed by the magnitude of \rd's opposing per-subunit driving, $\mu_1/n_1$. Consequently, \rd\ increasingly follows the direction of its own driving while \ru\ continues to rotate in the direction of \muh. As a result, communication between the two rotors breaks down: each rotor, on average, rotates in the direction of its own driving.

\section{Scaling driving}
Under a scaling force, since $\mu_{\rm o/1}/n_{\rm o/1} = \mu_{\rm o/1}^{\rm sub}$ is held constant for each rotor, the identity of the upstream and downstream rotor is fixed and does not change with subunit number. Without loss of generality, we use $\mu_{\rm o}^{\rm sub} > -\mu_1^{\rm sub}$, so \ru\ is always the upstream rotor. 

Increasing \nh\ increases the number of periods of the gear-like coupling potential over one revolution of \ru, so that \ru\ engages more frequently through the coupling with \rd. Under tight and intermediate coupling, this frequent engagement increases output power with \nh\ over much of the parameter range (Fig.~\ref{fig:P1_vs_n0_mod4}). This increase, however, does not hold universally. At large per-subunit driving $\mu_{\rm o}^{\rm sub}$ in \ru, as \nh\ increases the output power rises to a maximum and then decreases, becoming negative at sufficiently large \nh\ while input power remains positive (Fig.~\ref{fig:P0_vs_n0_mod4}). This is disruption: the same breakdown of energy transduction seen under constant driving (Sec.~\ref{subsubsec:disruption_gears}), beyond which, each rotor rotates on average in the direction of its own driving. 

\section{Conclusions}
In this study, we quantify how the output power of two gear-like coupled rotors varies with the number of subunits (and thus the symmetry) of the input rotor. The results of this study provide a theoretical perspective on two key interfaces in the BFM: the C-ring--MS-ring interaction (motivating constant driving) and the stator--C-ring interaction (motivating scaling driving). 

For the C-ring--MS-ring interaction, it is plausible that the C-ring drives the MS-ring through a constant driving force, since the driving originates from the stators rather than from the C-ring and thus is independent of the C-ring's subunit count. The MS-ring has no direct external driving ($\mu_{\scriptscriptstyle\mathrm{MS}} = 0$), and the two rings are strongly bound~\cite{kawamoto_native_2021}. Under constant driving and gear-like coupling, our results show that output power is maximized at symmetry match when the coupling is tight and the upstream driving is strong. Under weaker coupling, however, the power maximum at symmetry match turns into a local minimum. This finding provides a possible physical rationale for the observed distribution of C-ring subunit numbers in \emph{Salmonella}'s BFM, which ranges from 32 to 36, with 34-fold symmetry (matching the MS-ring) most common~\cite{kawamoto_native_2021}: a C-ring subunit number near that of the MS-ring may be selected to maximize the rotational speed of the MS-ring and, consequently, the flagellum. 

Differences in the sizes, molecular composition, and local environments of the C-ring (upstream) and MS-ring (downstream) may give rise to different friction coefficients. As a robustness test, we therefore additionally considered unequal friction coefficients together with a passive \rd\ (MS-ring) ($\mu_{\scriptscriptstyle\mathrm{MS}} = 0$); under strong coupling, at sufficiently strong driving, the rotational response again is maximized near symmetry match (Sec.~\ref{SM:C_S_ring}). 

Under scaling driving, strong coupling, and comparable per-subunit drivings on the two rotors, we find that output power increases monotonically with the upstream subunit number. Assuming that each additional stator contributes approximately additively to the total driving on the C-ring~\cite{nord_speed_2017}, increasing the number of stators can be mapped onto increasing the upstream subunit number under our scaling-driving scheme. This suggests that the C-ring's output power benefits from accommodating more stators, which is consistent with evidence under high load that the rotational speed of the BFM increases with the number of active stators~\cite{nakamura_flagella-driven_2019}. 

\begin{acknowledgments}
This work was supported by an SFU Graduate Dean's Entrance Scholarship (SI), a Natural Sciences and Engineering Research Council of Canada (NSERC) Discovery Grant and Discovery Accelerator Supplement RGPIN-2020-04950, an NSERC Alliance International Collaboration Grant ALLRP-2023-585940, and a Tier-II Canada Research Chair CRC-2020-00098 (all DAS). This research was enabled in part by support provided by BC DRI Group and the Digital Research Alliance of Canada (www.alliancecan.ca).
\end{acknowledgments}
\vspace{2ex}
{\bf Data Availability:} the data that support the findings of this article are openly available~\cite{iranbakhsh2026dataset}.

\FloatBarrier
\nocite{reimann_giant_2001,risken_fokker-planck_1996}
\bibliographystyle{unsrt}
\bibliography{test_python_done,github}

\newpage

\renewcommand{\thesection}{S\arabic{section}}
\renewcommand{\thesubsection}{\thesection.\arabic{subsection}}
\renewcommand{\thesubsubsection}{\thesubsection.\arabic{subsubsection}}

\makeatletter
\renewcommand{\p@subsection}{}
\renewcommand{\p@subsubsection}{}
\makeatother

\renewcommand{\theequation}{S\arabic{equation}}
\renewcommand{\thefigure}{S\arabic{figure}}
\renewcommand{\thetable}{S\arabic{table}}

\setcounter{section}{0}
\setcounter{page}{1}
\setcounter{figure}{0}

\onecolumngrid

\begin{center}
    \textbf{\large Supplemental Material for
    ``Two faces of symmetry match in gear-like coupled rotary motors''}
\end{center}

\section{Tight-coupling consequences for rotors under constant driving}
\label{app:tight_barriers}
Here, we analyze the tight-coupling consequences for a coupled system of two rotors with nonzero individual rotor barrier heights ($\beta E_{\rm o},\ \beta E_1>0$). The total energy landscape is
\begin{align}
    V(\theta_{\rm o},\theta_1)
    =
    -\frac{E_{\rm o}}{2}\cos(n_{\rm o}\theta_{\rm o})
    -\frac{E_1}{2}\cos(n_1\theta_1)
    -\frac{E_{\rm c}}{2}
    \cos(n_{\rm o}\theta_{\rm o}-n_1\theta_1).
    \label{eq:full_potential_tight_supp}
\end{align}
In the tight-coupling limit, the gear-like coupling constrains the weighted angular coordinates according to
\begin{equation}
    \phi  = \phi_{\rm o} = \phi_1,
    \label{eq:tight_constraint_supp}
\end{equation}
where $\phi_{\rm o} \equiv n_{\rm o}\theta_{\rm o}$ and $ \phi_1 \equiv n_1\theta_1$. The individual angular coordinates along the tightly coupled trajectory are therefore
\begin{subequations}
\begin{align}
    \theta_{\rm o} &=\frac{\phi}{n_{\rm o}} \\
    \theta_1 &= \frac{\phi}{n_1}.
\end{align}
\label{eq:phi_to_theta}
\end{subequations}
Substituting these relations into the individual rotor potentials gives
\begin{subequations}
\begin{align}
    V_{\rm o}(\phi) &= -\frac{E_{\rm o}}{2}\cos\phi\\
    V_1(\phi) &= -\frac{E_1}{2}\cos\phi.
\end{align}
\end{subequations}
Thus, along the tightly coupled trajectory, the two individual periodic landscapes add, and the coupling potential itself is constant. Therefore, the total energy landscape becomes
\begin{equation}
    V(\phi)=-\frac{E_{\rm o}+E_1}{2}\cos\phi-\frac{E_{\rm c}}{2}.
    \label{eq:locked_periodic_potential_supp}
\end{equation}

For an infinitesimal displacement $\mathrm{d}\phi$, the work performed by the external driving forces is
\begin{align}
    \delta W &= \mu_{\rm o}\,\mathrm{d}\theta_{\rm o} + \mu_1\,\mathrm{d}\theta_1 \nonumber\\
    &=
    \left(
    \frac{\mu_{\rm o}}{n_{\rm o}}
    +
    \frac{\mu_1}{n_1}
    \right)\mathrm{d}\phi.
\end{align}
The generalized driving force conjugate to $\phi$ is therefore the total driving introduced in~\eqref{eq:total_driving_gears}:
\begin{equation}
    \mu_{\rm tot}= \frac{\mu_{\rm o}}{n_{\rm o}}
    + \frac{\mu_1}{n_1}.
    \label{eq:mu_eff_tight_barriers_supp}
\end{equation}

The instantaneous viscous dissipation rate of the two rotors is
\begin{align}
    \mathcal{P}_{\rm visc}
    &=
    \gamma_{\rm o} \, (\partial_t\theta_{\rm o})^{\,2}
    +
    \gamma_1 \, (\partial_t\theta_1)^{\,2}\nonumber\\
    &= \left(
    \frac{\gamma_{\rm o}}{n_{\rm o}^{2}}
    +
    \frac{\gamma_1}{n_1^{2}}
    \right)(\partial_t\phi)^{\,2}.
\end{align}
This identifies a total friction coefficient
\begin{equation}
    \gamma_{\rm tot}
    =
    \frac{\gamma_{\rm o}}{n_{\rm o}^{2}}
    +
    \frac{\gamma_1}{n_1^{2}}.
    \label{eq:gamma_eff_tight_supp}
\end{equation}
For the equal friction coefficients used in the numerical simulations ($\gamma_{\rm o}=\gamma_1=\gamma$),
\begin{equation}
    \gamma_{\rm tot}
    =
    \gamma
    \left(
    \frac{1}{n_{\rm o}^{2}}
    +
    \frac{1}{n_1^{2}}
    \right).
    \label{eq:gamma_eff_equal_supp}
\end{equation}

While our numerical simulations use the Fokker–Planck equation~\eqref{eq:FPE} to describe the evolution of the probability distribution, we here use the equivalent Langevin description, which describes individual stochastic trajectories. Equations~\eqref{eq:locked_periodic_potential_supp}, \eqref{eq:mu_eff_tight_barriers_supp}, and \eqref{eq:gamma_eff_equal_supp} reduce the original two-dimensional tightly coupled system to the one-dimensional Langevin equation
\begin{subequations}
\begin{align}
    \gamma_{\rm tot}\partial_t\phi
    &=
    -\frac{\mathrm{d}V(\phi)}{\mathrm{d}\phi}
    +
    \mu_{\rm tot}
    +
    \eta_{\rm tot}
    \label{eq:phi_langevin_supp} \\
    &=
    -\frac{E_{\rm o}+E_1}{2}\sin\phi
    +
    \mu_{\rm tot}
    +
    \eta_{\rm tot}.
    \label{eq:phi_langevin_explicit_supp}
    \end{align}
\end{subequations}
The total noise $\eta_{\rm tot}$ satisfies the fluctuation--dissipation relation
\begin{equation}
    \left\langle
    \eta_{\rm tot}(t) \, \eta_{\rm tot}(t')
    \right\rangle
    =
    2\gamma_{\rm tot} \, k_{\rm B}T
    \, \delta(t-t').
    \label{eq:eta_eff_supp}
\end{equation}

The reduced dynamics in~\eqref{eq:phi_langevin_explicit_supp} have the same form as those of a single overdamped rotary motor on a tilted periodic potential. We therefore apply the standard steady-state velocity expression used previously for a single rotary motor~\cite{reimann_giant_2001}:
\begin{equation}
    \left\langle\partial_t\phi\right\rangle
    =
    \frac{
    1-\exp\{-2\pi\beta\mu_{\rm tot}\}
    }{
    \displaystyle
    \frac{\beta\gamma_{\rm tot}}{2\pi}
    \int_0^{2\pi}\mathrm{d}\phi\,
    \int_0^{2\pi}\mathrm{d}\phi'\,
    \exp\left\{
    \beta\left[
    V(\phi)-V(\phi-\phi')
    -\mu_{\rm tot}\phi'
    \right]
    \right\}
    }.
    \label{eq:phi_velocity_exact_supp}
\end{equation}
For fixed $\mu_{\rm o}$, $\mu_1$, $n_1$, $E_{\rm o}$, and $E_1$, Eq.~\eqref{eq:phi_velocity_exact_supp} provides the tight-coupling prediction for the output power as a function of upstream subunit number \nh:
\begin{equation}
    \mathcal{P}_1^{\rm tight}(n_{\rm o})
    =
    -\frac{\mu_1}{n_1}
    \left\langle\partial_t\phi\right\rangle_{n_{\rm o}}.
    \label{eq:P1_n0_tight_supp}
\end{equation}
Because the physical subunit number is discrete, the final theoretical prediction is obtained by evaluating $\mathcal{P}_1^{\rm tight}$ at each integer $n_{\rm o}$ considered in the simulations and selecting
\begin{equation}
    n_{\rm o}^*
    =
    \underset{n_{\rm o}\in\mathcal{N}}{\operatorname{arg\,max}}\ 
    \mathcal{P}_1^{\rm tight}(n_{\rm o}) \ ,
    \label{eq:n0_star_tight_barriers_supp}
\end{equation}
where $\mathcal{N}$ denotes the set of upstream subunit numbers considered. No fitting parameters are used.

\subsection{The barrierless result}
\label{app:tight_barrierless}
Without barriers ($E_{\rm o}=0=E_1$), the periodic individual potentials vanish. Thus, \eqref{eq:phi_langevin_supp} describes a freely driven Brownian particle with average velocity
\begin{equation}
    \left\langle\partial_t\phi\right\rangle
    =
    \frac{\mu_{\rm tot}}{\gamma_{\rm tot}}.
\end{equation}
Using~\eqref{eq:total_driving_gears} and~\eqref{eq:gamma_eff_equal_supp}, with $\kappa=n_1/n_{\rm o}$, gives
\begin{equation}
    \left\langle\partial_t\theta_1\right\rangle
    =
    \frac{\kappa\mu_{\rm o}+\mu_1}
    {\gamma(1+\kappa^2)}.
    \label{eq:recover_barrierless_velocity_supp}
\end{equation}
Maximizing \eqref{eq:recover_barrierless_velocity_supp} with respect to $\kappa$ gives
\begin{equation}
    \mu_{\rm o}\kappa^2 + 2\mu_1\kappa-\mu_{\rm o}=0 \ ,
\end{equation}
whose physically relevant root is
\begin{equation}
    \kappa^{*}=-\frac{\mu_1}{\mu_{\rm o}}+\sqrt{1+\left(\frac{\mu_1}{\mu_{\rm o}}\right)^2}.
    \label{eq:kappa_star_app}
\end{equation}

Two features of \eqref{eq:kappa_star_app} are worth noting. First, $\kappa^{*}$ depends only on the ratio $\mu_1/\mu_{\rm o}$. Second, in the limit $\mu_{\rm o} \gg -\mu_1$, expanding the square root to second order in $\mu_1/\mu_{\rm o}$ gives
\begin{subequations}
    \begin{align}
    \kappa^{*}
    &\approx 1 - \frac{\mu_1}{\mu_{\rm o}} + \frac{1}{2}\left(\frac{\mu_1}{\mu_{\rm o}}\right)^2 \\
    &
    \to 1 \, .
    \label{eq:kappa_star_limit}
\end{align}
\end{subequations}
Since $\kappa = n_1/n_{\rm o}$ for gear-like coupling, $\kappa^* \to 1$ corresponds to $n_{\rm o}^* \to n_1$: the barrierless optimum approaches symmetry match as the driving on \ru\ dominates.
\begin{figure*}[t!]
    \centering
    \includegraphics[width=\textwidth]{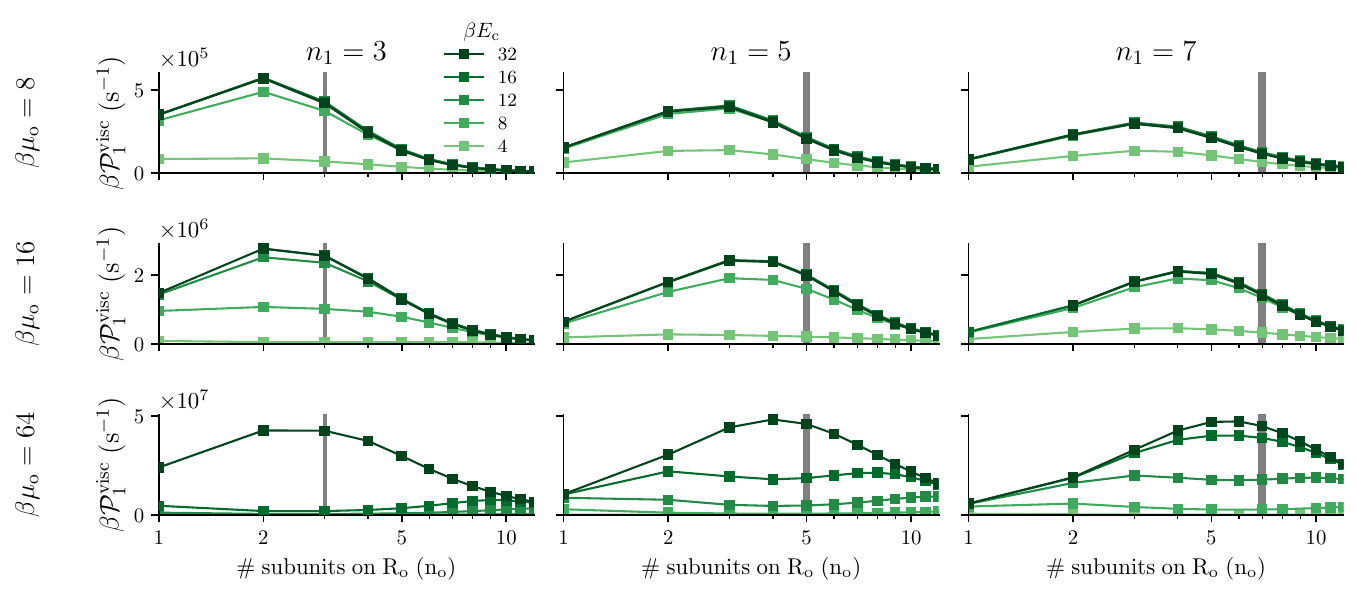}
    \caption{Viscous dissipation $\beta\mathcal{P}_1^{\rm visc}$ associated with the rotation of the passive downstream rotor, as a function of the number $n_{\rm o}$ of upstream-rotor subunits for different coupling strengths $\beta E_{\rm c}$. Rows each have fixed upstream driving $\beta\mu_{\rm o}$ on the upstream rotor ranging from $8$ to $64$; columns each have fixed downstream subunit number $n_1$ ranging from $3$ to $7$. Gray vertical solid lines: symmetry match ($n_{\rm o}=n_1$). $\beta\mu_1=0$, $\gamma_{\rm o}=1000$, $\gamma_1=1500$, and $\beta E_{\rm o}=\beta E_1=2$ throughout.}
    \label{fig:CS_ring_supp}
\end{figure*}
\section{C-ring--MS-ring-inspired simulations}
\label{SM:C_S_ring}
To test whether the symmetry-dependent behavior found for coupled rotors persists under conditions more closely mirroring the C-ring--MS-ring interaction of the BFM, we additionally consider a driven upstream rotor coupled to a passive downstream rotor. We set the downstream driving to zero ($\mu_1=0$) and use unequal friction coefficients, $\gamma_{\rm o}=1000$ and $\gamma_1=1500$. 

Because the downstream rotor is passive, we characterize its rotational response using the viscous dissipation associated with its rotation,
\begin{equation}
    \mathcal{P}_1^{\rm visc} = \gamma_1(2\pi J_1)^2. 
    \label{eq:CS_viscous_dissipation}
\end{equation}

Figure~\ref{fig:CS_ring_supp} shows that at strong coupling, increasing the upstream driving $\mu_{\rm o}$ shifts the maximum downstream rotational response toward symmetry match, reproducing the qualitative progression observed for the coupled rotors in Fig.~\ref{fig:P1_vs_n0_mod3}. The correspondence is weaker at lower coupling strengths, consistent with the importance of maintaining coordinated motion between the two rotors. These results show that the symmetry-dependent behavior identified in the general model can persist when the downstream rotor is passive and the two rotors have unequal friction coefficients.

\section{Barrierless rotors}
The following set of Langevin equations describes barrierless rotors coupled through a gear-like coupling: 
\begin{subequations}  
\begin{align}
\gamma_{\rm o} \partial_t\theta_{\rm o} + 
n_{\rm o}\frac{E_{\rm c}}{2} 
\sin(n_{\rm o}\theta_{\rm o} - n_1\theta_1) - \mu_{\rm o} &= \eta_{{\rm o}} \, \label{eq:theta0_barrierless_coupled_gear}
\\
\gamma_1 \partial_t\theta_1 - n_1 \frac{E_{\rm c}}{2} \sin(n_{\rm o}\theta_{\rm o} - n_1\theta_1) - \mu_{\rm 1} &= \eta_{1}. 
\label{eq:theta1_barrierless_coupled_gear}
\end{align}
\end{subequations}
Using a change of variables
\begin{subequations}
\begin{align}
\tilde{\phi} &\equiv n_1\theta_{\rm o} + n_{\rm o}\theta_1\\ 
\Delta\phi &\equiv n_{\rm o}\theta_{\rm o} - n_1\theta_1 , \label{eq:relative_coordinate}
\end{align}
\end{subequations}
the individual orientations are
\begin{subequations}
\begin{align}
\theta_{\rm o} &= \frac{n_1 \tilde{\phi} +n_{\rm o}\Delta\phi}{n_{\rm o}^2 + n_1^2} \\
\theta_1 &= \frac{n_{\rm o} \tilde{\phi} -n_1\Delta\phi}{n_{\rm o}^2 + n_1^2}.
\end{align}
\end{subequations}

Assuming equal friction coefficients $\gamma_{\rm o} = \gamma_1 = \gamma$, the set of Langevin equations~\eqref{eq:theta0_barrierless_coupled_gear}--\eqref{eq:theta1_barrierless_coupled_gear} become uncoupled in these new variables:
\begin{subequations}
\begin{align}
\gamma \, \partial_t \tilde{\phi} -  (n_1\mu_{\rm o} + n_{\rm o}\mu_{\rm 1}) &= \eta_{\tilde{\phi}}\label{eq:bar_theta} 
\\
\gamma \, \partial_t \Delta \phi + \left(n_{\rm o}^2 +n_1^2\right)\frac{E_{\rm c}}{2} \sin(\Delta \phi) - (n_{\rm o}\mu_{\rm o} - n_1\mu_{\rm 1}) &= \eta_{\Delta \phi}, \label{eq:delta_theta}
\end{align}
\end{subequations}
where
\begin{subequations}
\begin{align}
    \eta_{\tilde{\phi}} &\equiv n_1 \eta_{\rm o} + n_{\rm o} \eta_{1} \\
    \eta_{\Delta\phi} &\equiv n_{\rm o} \eta_{\rm o} - n_1 \eta_{1}.
\end{align}
\end{subequations}
The noise term $\eta_{\tilde{\phi}}$ has autocorrelation function
\begin{subequations}
\begin{align}
\langle \eta_{\tilde{\phi}}(t) \, \eta_{\tilde{\phi}}(t') \rangle
&= \Bigl\langle 
\bigl[n_1\eta_{\rm o}(t) + n_{\rm o}\eta_{1}(t)\bigr] \,
\bigl[n_1\eta_{\rm o}(t') + n_{\rm o}\eta_{1}(t')\bigr]
\Bigr\rangle \\[6pt]
&= \Bigl\langle 
n_1^2\eta_{\rm o}(t)\eta_{\rm o}(t') 
+ n_1 n_{\rm o}\eta_{\rm o}(t)\eta_{1}(t') 
+ n_1 n_{\rm o}\eta_{1}(t)\eta_{\rm o}(t') 
+ n_{\rm o}^2 \eta_{1}(t)\eta_{1}(t') 
\Bigr\rangle \\[6pt]
&= 2 n_1^2\gamma \, k_{\rm B} T \delta(t - t') 
+ 0 + 0 
+ 2 n_{\rm o}^2 \gamma \, k_{\rm B} T \delta(t - t') \\[6pt]
&= 2 (n_{\rm o}^2 + n_1^2)\,
\gamma \, k_{\rm B} T \,\delta(t - t'),
\end{align}
\end{subequations}
where the cross terms yield zero since $\eta_{\rm o}$ and $\eta_1$ are independent stochastic noise terms. Similarly,
\begin{equation}
\langle \eta_{\Delta\phi}(t) \, \eta_{\Delta\phi}(t') \rangle = 2\left(n_{\rm o}^2 + n_1^2\right) \, \gamma \, k_{\rm B} T \, \delta(t - t') .
\end{equation}

Using the analysis introduced in~\cite{risken_fokker-planck_1996}, the effective energy landscape for the weighted separation $\Delta\phi\equiv n_{\rm o}\theta_{\rm o}-n_1\theta_1$ is
\begin{equation}
    V_{\Delta\phi}(\Delta\phi)
    \equiv
    -\left(n_{\rm o}^2+n_1^2\right)\frac{E_{\rm c}}{2}\cos(\Delta\phi).
    \label{eq:Veff_delta}
\end{equation}
With this definition, \eqref{eq:delta_theta} can be written as
\begin{equation}
    \gamma\, \partial_t\Delta\phi
    =
    -\frac{\partial V_{\Delta\phi}}{\partial\Delta\phi}
    + \mu_{\Delta\phi} + 
    \eta_{\Delta\phi},
\end{equation}
where the effective driving along the relative coordinate is
\begin{equation}
    \mu_{\rm \Delta\phi} \equiv 
    n_{\rm o}\mu_{\rm o}-n_1\mu_1.
    \label{eq:effective_driving_gear}
\end{equation}

The fluxes associated with the transformed coordinates are
\begin{subequations}
\begin{align}
    J_{\tilde{\phi}}
    &\equiv
    \frac{1}{2\pi}
    \left\langle\partial_t\tilde{\phi}\right\rangle\\
    J_{\Delta\phi}
    &\equiv
    \frac{1}{2\pi}
    \left\langle\partial_t\Delta\phi\right\rangle,
    \label{eq:J_delta_theta}
\end{align}
\end{subequations}
where using \eqref{eq:bar_theta} gives
\begin{equation}
    J_{\tilde{\phi}}
    =
    \frac{n_1\mu_{\rm o}+n_{\rm o}\mu_1}
    {2\pi\gamma}.
\end{equation}
$J_{\Delta\phi}$ quantifies sustained relative motion between the two rotors and therefore provides a measure of slippage. The resulting physical rotor fluxes are
\begin{subequations}
\begin{align}
J_{\rm o}
&=
\frac{n_1}{n_{\rm o}^2+n_1^2}J_{\tilde{\phi}}
+
\frac{n_{\rm o}}{n_{\rm o}^2+n_1^2}J_{\Delta\phi}
\label{eq:J0_gear_barrierless}
\\
J_1
&=
\frac{n_{\rm o}}{n_{\rm o}^2+n_1^2}J_{\tilde{\phi}}
-
\frac{n_1}{n_{\rm o}^2+n_1^2}J_{\Delta\phi}.
\label{eq:J1_gear_barrierless}
\end{align}
\end{subequations}

Since $V_{\Delta\phi}$ has period $2\pi$, the steady-state average velocity of the relative coordinate is
\begin{equation}
\left\langle\partial_t\Delta\phi\right\rangle
    =
    \frac{
    1-\exp\left\{
    -
    \frac{\beta}{n_{\rm o}^2+n_1^2}2\pi\mu_{\rm \Delta\phi}
    \right\}
    }{
    \displaystyle
    \frac{\beta}{n_{\rm o}^2+n_1^2}\frac{\gamma}{2\pi}
    \int_0^{2\pi}\mathrm{d}\Delta\phi
    \int_0^{2\pi}\mathrm{d}\Delta\phi'
    \exp\left\{
    \frac{\beta}{n_{\rm o}^2+n_1^2}
    \left[
    V_{\Delta\phi}(\Delta\phi) - V_{\Delta\phi}(\Delta\phi-\Delta\phi')
    -
    \mu_{\rm \Delta\phi}
    \Delta\phi'
    \right]
    \right\}
    }.
    \label{eq:delta_velocity_exact}
\end{equation}

\subsection{Finite-coupling locking and slippage at strong driving}
\label{app:barrierless_locking}

The relative coordinate $\Delta\phi$ [Eq.~\eqref{eq:relative_coordinate}] provides a direct measure of relative motion between the two rotors. When the rotors remain locked, $\Delta\phi$ fluctuates about a minimum of the coupling potential; sustained motion in $\Delta\phi$ instead corresponds to one rotor advancing relative to the other and therefore to slippage.

Rearranging~\eqref{eq:delta_theta} gives
\begin{equation}
\gamma \, \partial_t \Delta\phi = - \left(n_{\rm o}^2+n_1^2\right) \frac{E_{\rm c}}{2} \sin(\Delta\phi) + \left( n_{\rm o}\mu_{\rm o} - n_1\mu_1 \right) + \eta_{\Delta\phi}.
\label{eq:delta_theta_locking}
\end{equation}
The first term on the right-hand side is the restoring effect of the gear-like coupling, which tends to maintain a fixed relative position of the two rotors. The second term drives relative motion. Since $\mu_{\rm o}>0$ and $\mu_1<0$ in the setup considered here, the two driving forces act together to increase the tendency for relative motion. Ignoring thermal fluctuations, a locked solution can exist only if the maximum restoring effect by the coupling is at least as large as the driving on the relative coordinate. From~\eqref{eq:delta_theta_locking}, this requires
\begin{equation}
    E_{\rm c}\geq E_{\rm c}^{\rm lock}=2\frac{n_{\rm o}\mu_{\rm o}-n_1\mu_1}{n_{\rm o}^2+n_1^2}.
    \label{eq:barrierless_locking_condition}
\end{equation}
This therefore gives a deterministic barrierless estimate of the coupling strength required to maintain locked motion. It is not an exact locking threshold for the full system with individual rotor barriers and thermal fluctuations, but it provides a simple measure of how the tendency toward slippage varies with symmetry.
\begin{figure*}[t]
    \centering
    \includegraphics[width=\textwidth]{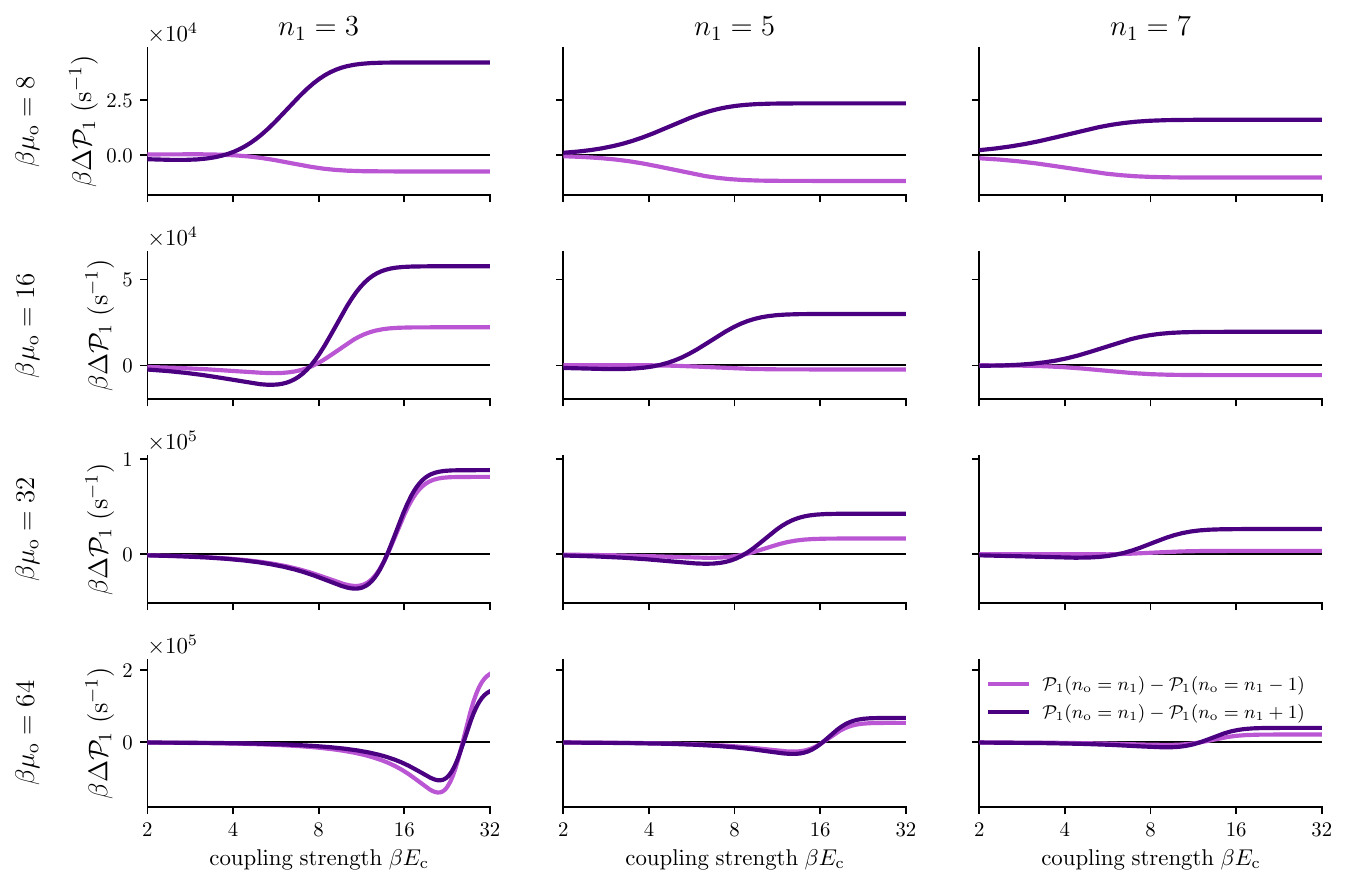}
    \caption{Output-power differences $\Delta \mathcal{P}_1$ between symmetry match and the nearest symmetry mismatches, calculated from the exact steady-state solution for barrierless rotors~\eqref{eq:delta_velocity_exact}. Pink curves show $\mathcal{P}_1(n_{\rm o}=n_1)-\mathcal{P}_1(n_{\rm o}=n_1-1)$, and purple curves show $\mathcal{P}_1(n_{\rm o}=n_1)-\mathcal{P}_1(n_{\rm o}=n_1+1)$. Rows each have fixed driving force $\beta\mu_{\rm o}$ on \ru\ ranging from 8 to 64; columns each have fixed number $n_1=n_{\rm o}$ ranging from 3 to 7. $\beta\mu_1=-2$ throughout.}
    \label{fig:barrierless_power_difference}
\end{figure*}

At fixed $\mu_{\rm o}$, $\mu_1$, and $n_1$, the right-hand side of~\eqref{eq:barrierless_locking_condition} is non-monotonic in $n_{\rm o}$. The value of $n_{\rm o}$ at which the required coupling is largest follows from
\begin{equation}
    \frac{\partial}{\partial n_{\rm o}}
    \left[\frac{n_{\rm o}\mu_{\rm o}-n_1\mu_1}{n_{\rm o}^2+n_1^2}\right]=0,
\end{equation}
which gives
\begin{equation}
    \mu_{\rm o}n_{\rm o}^2-2\mu_1 n_1n_{\rm o}-\mu_{\rm o} n_1^2=0.
\end{equation}
The positive solution is
\begin{equation}
    n_{\rm o}^{\rm lock} = n_1 \left(\frac{\mu_1}{\mu_{\rm o}}+ \sqrt{1+\left(\frac{\mu_1}{\mu_{\rm o}}\right)^2}\ \ \right) \ .
    \label{eq:n0_hardest_lock}
\end{equation}
Defining $\kappa^{\rm lock} \equiv n_1/n_{\rm o}^{\rm lock}$:
\begin{equation}
     \kappa^{\rm lock} = -\frac{\mu_1}{\mu_{\rm o}}+ \sqrt{1+\left(\frac{\mu_1}{\mu_{\rm o}}\right)^2}\ \ .
    \label{eq:kappa_hardest_lock}
\end{equation}
As the upstream driving becomes large compared to the opposing driving ($\mu_{\rm o}\gg-\mu_1$), $\kappa^{\rm lock} \rightarrow 1$; thus, the \nh\ for which locking requires the strongest coupling strength approaches \na\ ($n_{\rm o}^{\rm lock}\rightarrow n_1$). Hence, in the strong-driving limit, symmetry match gives particular susceptibility to relative motion. 

This result also connects directly to the barrierless tight-coupling gear ratio that maximizes power~\eqref{eq:kappa_star_limit}, which also gives $n_{\rm o}^*\rightarrow n_1$. Thus, for barrierless rotors, the gear ratio that maximizes output power when the rotors are tightly locked is also the ratio requiring the largest coupling strength to maintain that locking. At tight coupling, this ratio therefore produces the tight-coupling power maximum, whereas at weaker couplings and with sufficiently strong driving, increased relative motion near the same ratio can turn the maximum into a local minimum.

To directly examine whether finite coupling can turn the symmetry-match power maximum into a local minimum, we evaluate the exact barrierless solution~\eqref{eq:delta_velocity_exact} at $n_{\rm o}=n_1$ and its nearest neighboring symmetries $n_{\rm o}=n_1\pm1$ (Fig.~\ref{fig:barrierless_power_difference}). At sufficiently strong upstream driving, both power differences become negative at intermediate coupling strengths, indicating a local minimum at symmetry match. Increasing the coupling strength makes both differences positive, recovering a local maximum. Thus, the peak-to-dip inversion observed in the finite-barrier simulations also occurs in the barrierless model.

We emphasize that Eqs.~\eqref{eq:barrierless_locking_condition}--\eqref{eq:n0_hardest_lock} are derived for barrierless rotors. The numerical results in the main text are for positive individual-rotor barrier heights and thermal fluctuations, for which the transition from locked to slipping motion is expected to be a smooth crossover rather than the sharp condition in~\eqref{eq:barrierless_locking_condition}.
\begin{figure*}[t!]
    \centering
    \includegraphics[width=\textwidth]{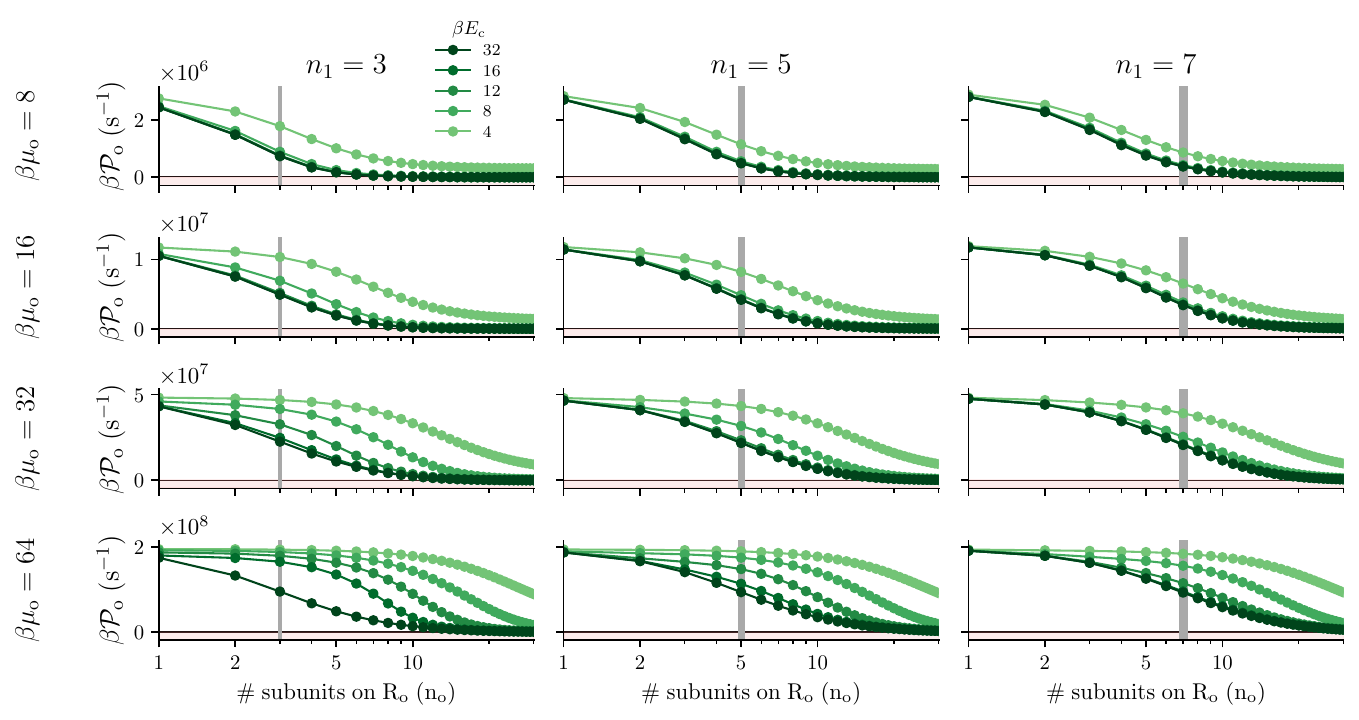}
    \caption{Input power as a function of the number \nh\ of \ru\ subunits, for different coupling strengths $\beta E_{\rm c}$ under constant driving. Rows each have fixed driving force $\beta\mu_{\rm o}$ on \ru\ ranging from 8 to 64; columns each have fixed number $n_1$ of \rd\ subunits ranging from 3 to 7. Gray vertical solid lines: symmetry match ($n_1 = n_{\rm o}$). $\beta \mu_1 = -2$ and $\beta E_{\rm o} = \beta E_1 = 2$ throughout.}
    \label{fig:P0_vs_n0_mod3}
\end{figure*}

\section{Disruption}
\label{app:disruption}
\begin{figure*}[htbp]
    \centering
    \includegraphics[width=\textwidth]{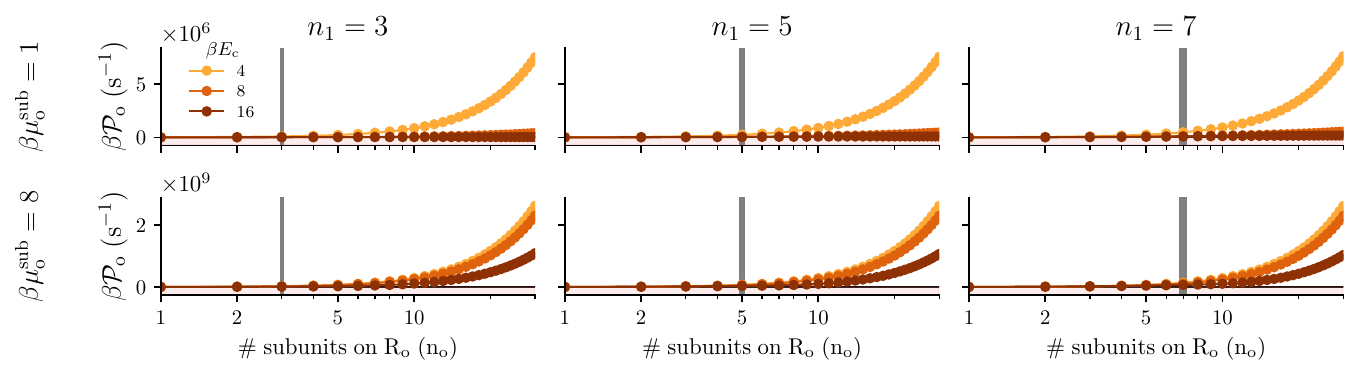}
    \caption{Input power under scaling driving as a function of the number \nh\ of \ru\ subunits, for different coupling strengths $\beta E_{\rm c}$. Rows have fixed per-subunit driving force $\beta\mu_{\rm o}^{\rm sub}$ on \ru\ (1 on the top and 8 on the bottom row) and columns have fixed number $n_1$ of subunits of \rd\ ranging from 3 to 7. Gray vertical solid lines show symmetry match $n_1 = n_{\rm o}$. $\beta \mu_1^{\rm sub} = -2/3$ and $\beta E_{\rm o} = \beta E_1 = 2$ throughout.}
    \label{fig:P0_vs_n0_mod4}
\end{figure*}

Figures~\ref{fig:P0_vs_n0_mod3} and~\ref{fig:P0_vs_n0_mod4} show input power \ph\ as a function of \nh, corresponding to the output powers in Fig.~\ref{fig:P1_vs_n0_mod3} and Fig.~\ref{fig:P1_vs_n0_mod4}, respectively. Disruption occurs when positive input no longer yields positive output as a result of the increase in the number \nh\ of subunits on the upstream system while fixing the coupling. For instance, in the top-left panel of Fig.~\ref{fig:P1_vs_n0_mod3}, the $\beta E_{\rm c} = 4$ output-power curve crosses zero at $n_{\rm o} = 5$, while in the top-left panel of Fig.~\ref{fig:P0_vs_n0_mod3}, the corresponding input power remains positive. 
\FloatBarrier

\end{document}